\scrollmode
\documentclass[english,12pt]{article}
          \usepackage{babel}
          \usepackage[T1]{fontenc}
          \usepackage[latin2]{inputenc}
          \usepackage{pslatex}
\input epsf

\begin{document}

\voffset -0.5in
\title{The recurrent nova U Scorpii -- an evolutionary considerations}

\author{Marek J. Sarna$\rm ^{1}$, Ene Ergma$\rm ^{2}$ and
Jelena Ger\v{s}kevit\v{s}$\rm^{2,1}$}
\date{$\rm  ^1~$ N. Copernicus Astronomical Center,
       Polish Academy of Sciences,
       ul. Bartycka 18, 00-716 Warsaw, Poland \\
       e-mail: sarna$@$camk.edu.pl; jelena@camk.edu.pl \\
$\rm  ^2~$ Physics Department, Tartu University, \"Ulikooli 18,
50510 Tartu, Estonia \\
       e-mail: ene$@$physic.ut.ee; jelen\_a@physic.ut.ee}

\maketitle

\begin{abstract}
We perform evolutionary calculations of binary stars  to find
progenitors of systems with parameters similar to the recurrent
nova U Sco.  We show that a U Sco-type system may be formed
starting with an initial binary system which has a low-mass
carbon-oxygen white dwarf as an accretor. Since the evolutionary
stage of the secondary is not well known, we calculate sequences
with hydrogen-rich and helium-rich secondaries. The evolution of
the binary may be divided into several observable stages as:
classical nova, supersoft X-ray source with  stable hydrogen
shell burning, or strong wind phase.  It culminates in the
formation of a massive white dwarf near the Chandrasekhar mass
limit. We follow the chemical evolution of the secondary as well
as of the matter lost from the system, and we show that observed
$^{12}$C/$^{13}$C and N/C ratios may give some information about
the nature of the binary.
\end{abstract}

\section{Introduction}

Recurrent novae are a small class of objects which bear many similarities
to other cataclysmic variable systems. They experience recurrent outbursts
at intervals of 10--80 yrs.

 Recurrent novae have been proposed as possible progenitors of type Ia
supernovae (Starrfield, Sparks \& Truran 1985). Although they
experience recurrent outbursts it is thought that the white dwarf
mass would grow to finally exceed Chandrasekhar mass limit.

Webbink et al. (1987) discussed the nature of the recurrent novae,
and they concluded that according to outburst mechanisms there are
two subclasses of these systems: (a) powered by thermonuclear
runaway on the surface of the white dwarf (e.g. U Sco),
and (b) powered by the transfer of a burst of matter from the red giant
to the main-sequence companion.

U Sco is one of the best observed recurrent novae. Historically,
its outbursts were observed in 1863, 1906, 1917, 1936, 1979, 1987
and in 1999. Schaefer (2004) discovered on Harvard College
Observatory archival photographs unknown eruption of the recurrent
nova U Sco in 1917. Schaefer notes that U Sco has a fairly
constant recurrent cycle of 8--12 yrs, with about 25\% of the
outbursts being missed due to proximity to the sun (including
potential missed outbursts around 1926 and 1957). Determinations
of the system visual luminosity at maximum and minimum indicate a
range $\Delta m_V \sim$ 9. Schaefer (1990) and Schaefer \&
Ringwald (1995) observed eclipses of U Sco in the quiescent phase,
and determined the orbital period $P_{\rm orb}$=1.23056 d.

For the 1979 outburst, ejecta abundances have been estimated from
optical and UV studies by Williams et al. (1981) and Barlow et al.
(1981). They derived extremely helium rich ejecta He/H$\sim$2 (by
number), while the CNO abundance was solar with an enhanced N/C
ratio. From analysis of the 1999 outburst, Anupama \& Dewangan
(2000) obtained an average helium abundance by number of
He/H$\sim$0.4$\pm$0.06, while Iijima (2002) 0.16$\pm$0.02.  The
estimated mass of the ejected shell for 1979 and 1999 outbursts is
$\rm \sim 10^{-7} ~M_\odot$ (Williams et al. 1981; Anupama \&
Dewangan 2000). Hachisu et al. (2000) and Matsumoto, Kato \&
Hachisu (2003) estimated the envelope mass at the optical maximum
as being $\rm \sim 3 \times 10^{-6} ~M_\odot $. If this is the
present case, the mass accretion rate should be smaller than $\rm
\sim 2.5 \times 10^{-7} ~M_\odot ~yr^{-1} $ in the quiescence
between 1987 and 1999. Spectroscopically, U Sco shows very high
ejection velocities of (7.5--11)$\rm \times 10^3 ~km ~s^{-1}$
(Williams et al. 1981; Munari et al. 2000). Latest determinations
of the spectral type of the secondary indicate a K2 subgiant
(Anupama \& Dewangan 2000; Kahabka et al. 1999). According to
Kahabka et al. (1999) the distance to U Sco is about 14 kpc.

Recently, the spectroscopic orbit of U Sco was determined by
Thoroughgood et al. (2001).  The radial velocity
semiamplitudes for primary and secondary stars yield
$M_{\rm wd} = 1.55 \pm 0.24 ~{\rm M}_{\odot}$ for the white dwarf and
$M_2 = 0.88 \pm 0.17 {\rm M}_{\odot}$ for a secondary star.

According to a model proposed by Kato (1996), supersoft X-ray
emission should be observed about 10--60 days after the optical
outburst.  Indeed, BeppoSAX detected X-ray emission from U Sco at
0.2--20 keV just 19--20 days after the peak of its optical
outburst in February 1999 (Kahabka et al.  1999). The fact that U
Sco was detected as a supersoft X-ray source (SSS) is consistent
with steady hydrogen shell burning on the surface of its white
dwarf component.

In this paper we construct a grid of evolutionary sequences which
may lead to formation of U Sco-like systems. In Section 2 we
discuss the major phases of semidetached binary evolution. In
Section 3 the evolutionary code is briefly described. Section 4
contains the results of the calculations. A general discussion and
conclusions follow.

\section{The major modes of semidetached binary evolution}

We can identify three different modes of semidetached
evolution of a close binary system with white dwarf as accretor:

\noindent i) novae (nova outbursts);

\noindent ii) stable hydrogen shell burning on a compact
white dwarf; and

\noindent iii) a strong, optically thick wind powered by
steady hydrogen burning.

While calculating evolutionary models of binary stars, we must
take into account mass transfer and associated physical mechanisms
which lead to mass and angular momentum loss. We can express the
change in the total orbital angular momentum ($J$) of a binary
system as

\begin{equation}
\frac{\dot{J}}{J} = \left. \frac{\dot{J}}{J} \right|_{\rm MSW} +
\left. \frac{\dot{J}}{J} \right|_{\rm NOAML} + \left.
\frac{\dot{J}}{J} \right|_{\rm FWIND} + \left. \frac{\dot{J}}{J}
\right|_{\rm FAML}
\end{equation}

\noindent
where the terms on the right hand side are due to: magnetic
stellar wind braking (MSW); nova outburst angular momentum loss,
which describes the loss of angular momentum from the system due to
non-conservative evolution (NOAML); optically thick wind angular
momentum loss from the white dwarf during stable hydrogen shell
burning (FWIND); and frictional angular momentum loss in
a strong optically thick wind (FAML).

To account for magnetic stellar wind braking, we use the
Skumanich (1972) `law',

\begin{eqnarray}
\left. \frac{\dot{J}}{J} \right|_{\rm MSW} & = & - \left\{ \left( 5.0
\times 10^{-29} \right) \: \left( {\rm 2 \: \pi} \right)^\frac{10}{3}
\: {\rm G}^{-\frac{2}{3}} \right\} \nonumber \\
      &   & \mbox{} \left( \frac{k_{2}}{\lambda} \right)^{2} \:
\frac{M^{\frac{1}{3}}_{\rm tot} \: {R^{4}_{\rm sg}}}{M_{\rm wd}} \:
P^{-\frac{10}{3}}_{\rm orb}
\end{eqnarray}

\noindent
where $M_{\rm wd}$ and $M_{\rm tot}$ denote, respectively, the mass of the white dwarf
primary and the total mass of the system; $R_{\rm sg}$ is
the radius of the secondary star; $k_2 $ is the radius of gyration
of the secondary star, $k^2_2 = 0.1$ (Webbink 1976) and $\lambda $
characterizes the effectiveness of dynamo action in the
stellar convective zone. Values of $\lambda = 0.73$ (Skumanich
1972) or $\lambda = 1.78$ (Smith 1979) can be found in the
literature; however, we use a value of $\lambda = 1$ according to
Ergma \& Sarna (2000) model calculations (the best fit to the
observed orbital decay of the PSR B1744--24A).  We assume that MSW losses are
present in all three modes of evolution identified here.

\subsection{Novae}

In novae, angular momentum loss
accompanies mass loss due to the nova outbursts.

\subsubsection{Angular momentum loss associated with nova outbursts}

To take account of the angular momentum loss accompanying mass
loss in nova outbursts, we use a
formula based on that used to calculate angular momentum loss via
a stellar wind (Paczy\'nski \& Zi\'o\l kowski 1967; Zi\'o\l kowski
1985 and De Greve 1993),

\begin{equation}
\left. \frac{\dot{J}}{J} \right|_{\rm NOAML} = f_{1} \: f_{2} \:
\frac{M_{\rm wd} \: {\dot{M}}_{\rm sg}}{M_{\rm sg} \: M_{\rm tot}}, \:\:\:\:\: {\rm where}
\:\:\:\:\: \dot{M} = f_{1} \: {\dot{M}}_{\rm sg}.
\end{equation}

\noindent
Here, $f_{1}$ is the ratio of the mass ejected by the white dwarf to
that accreted by the white dwarf,

\begin{equation}
f_1 = {{\Delta M_{\rm ej}} \over {\Delta M_{\rm acc}}} ~ ,
\end{equation}

\noindent where $\Delta M_{\rm acc} $ and $\Delta M_{\rm ej} $ are
the amounts of matter transferred from the secondary to the white
dwarf and ejected in a nova outburst, during one nova cycle,
respectively. The parameter $f_{2}$ described the effectiveness of
angular momentum loss during mass transfer (Sarna \& De Greve
1994, 1996); $ M_{\rm sg} $ denotes the mass of the subgiant
secondary; ${\dot{M}}_{sg}$ is the rate of mass loss from the
secondary ($-{\dot{M}}_{\rm sg}$ is equivalent to the mass
transfer rate). We employ the code developed by Marks \& Sarna
(1998) (hereafter MS98), which utilizes the results of the
theoretical novae calculations  of Prialnik \& Kovetz (1995) and
Kovetz \& Prialnik (1997) (hereafter PK95 and KP97). Using this
data, we are able to compute the value of $f_1$, which we have
previously had to assume as free parameter (Sarna, Marks \& Smith
1996). By interpolation from the data of PK95 and KP97, at each
time step, we use the white dwarf mass and the mass transfer rate
to determine the nova characteristics: $f_1$, amplitude of the
outburst (A), recurrence period ($ P_{\rm rec} $), and chemical
composition of the ejected material (for more details see MS98).

We take $f_{2} = 1.0$, which is typical for a stellar wind. Livio
\& Pringle (1998) proposed a model in which the accreted angular
momentum is removed from the system during nova outbursts,
as previously suggested by Marks, Sarna \& Prialnik
(1997). Equation (3) is in quantitative agreement with these
estimations.

As discussed by MS98, angular momentum loss rates from
nova outbursts are comparable with MSW losses.
As an illustration, we can estimate the characteristic time scales
[$\tau = (\dot{J}/J)^{-1} $] for both mechanisms. We assume a
binary system in the same evolutionary stage as U Sco, with components of 0.94
and 1.3$~M_\odot $ and orbital period 1.23 d. The secondary has a
radius 2.22~${\rm R}_\odot$; $f_1 = 0.92$ and the mass transfer rate from
the secondary to white dwarf is $3.3 \times 10^{-8} ~{\rm M}_\odot
~{\rm yr}^{-1}$. We obtain $\tau_{\rm MSW} \sim 1.5 \times 10^{16}$ s and
$\tau_{\rm NOAML} \sim 1.7 \times 10^{15}$ s, i.e., angular momentum losses
associated with nova outbursts is about an order of magnitude greater than
those from the magnetic stellar wind.

\subsubsection{Accretion of material ejected during nova outbursts}

To calculate the re-accretion by the secondary of material ejected
during nova outbursts, we assume that the mass of the material
re-accreted ($M_{\rm re-acc}$) is proportional to the mass of the
material ejected by the white dwarf, such that

\begin{equation}
M_{\rm re-acc} = \left( \frac{{R_{\rm sg}}}{2 a} \right)^{2} \Delta M_{\rm ej}
\:\:\:\:\:\: .
\end{equation}

\noindent
The constant of proportionality is the ratio of the cross-section
area of the secondary star to the area of a sphere at radius $a$
from the white dwarf. We base this formula on the assumption that
nova ejections are spherically symmetric, with ejection velocities large
compared with the orbital velocity of the companion.

During nova outbursts, matter which is re-accreted by the
secondary changes significantly its chemical composition.

\subsection{Stable hydrogen shell burning}

We assume that during stable hydrogen shell burning, only angular
momentum losses due to the magnetic stellar wind will be important.

\subsection{Strong optically thick wind}

In this phase, two more processes change the total
angular momentum and mass of the system: a strong
optically thick wind, and frictional deposition of orbital energy
into that wind.

\subsubsection{Loss of angular momentum due to an optically thick wind}

If we consider a carbon-oxygen (C-O) white dwarf accreting matter from a
companion
with solar composition, there exists a critical accretion rate above
which the excess material is blown off in a strong wind.
Hachisu et al. (1996) show that because wind velocity is about
10 times higher than orbital velocity, the wind has the same
specific angular momentum as the white dwarf.  The corresponding loss rate is

\begin{equation}
\left. \frac{\dot{J}}{J} \right|_{\rm FWIND} = {q \over {(1 +q)}}
{{{\dot M}_{\rm wind}} \over M_{\rm wd}} ~ ,
\end{equation}

\noindent
where $q=M_{\rm sg}/M_{\rm wd}$ is the mass ratio and $ {\dot M}_{\rm wind} $
is the mass loss rate from the system. This mechanism will be very
effective during strong optically thick wind. Note also that
during the wind phase, the relatively small specific angular
momentum content of the wind tends to stabilize mass transfer
(Hachisu et al. 1996; Li \& van den Heuvel 1997).

\subsubsection{Frictional angular momentum loss}

During an optically thick wind phase, the secondary star effectively
orbits within the dense wind. Due to the frictional
deposition of orbital energy into the wind, the separation of the
components will be decreased.

Livio, Govarie \& Ritter (1991) estimated the change in the orbital angular
momentum brought about by frictional angular momentum loss,

\begin{equation}
\left. \frac{\dot{J}}{J} \right|_{\rm FAML} = \frac{1}{4}
\left( 1 + q \right) \: \frac{\left( 1 + U^{2}
\right)^{\frac{1}{2}}}{U}
\left( \frac{R_{\rm sg}}{a}\right)^{2} \: \frac{{\dot{M}}_{\rm wind}}{M_{\rm sg}} ,
\end{equation}

\noindent
where $U \equiv (v_{\rm exp}/v_{\rm orb})$ is the ratio of the expansion
velocity of the envelope at
the position of the secondary to the orbital velocity of the secondary
in the primary's frame of reference,
$a$ is the separation of the system, and ${\dot{M}}_{\rm wind}$ is the rate
of mass flow past the secondary (Warner 1995). Hachisu, Kato \& Nomoto (1996)
estimated that $v_{\rm exp} / v_{\rm orb} \sim 10$.

\subsubsection{Re-accretion of material from strong wind}

We define two critical mass accretion rates onto white dwarf. First (Warner
1995):

\begin{equation}
{\dot M}_{\rm cr,1} = 2.3 \times 10^{-7} (M_{\rm wd} - 0.19)^{3/2} ~
[{\rm M}_\odot ~{\rm yr}^{-1}] ~,
\end{equation}

\noindent
where $\dot M_{\rm cr,1}$ is the critical accretion rate above which
hydrogen shell burning is stable. Second (Nomoto, Nariai \&
Sugimoto 1979; Hachisu et al. 1996):

\begin{equation}
{\dot M}_{\rm cr,2} = 9.0 \times 10^{-7} (M_{\rm wd} - 0.5) ~
[{\rm M}_\odot ~{\rm yr}^{-1}] ~,
\end{equation}

\noindent
where ${\dot M}_{\rm cr,2}$ is the critical accretion rate above which a
strong wind occurs. In the latter case, the hydrogen is burning
into helium at a rate close to ${\dot M}_{\rm cr,2} $, with the excess
material being blown off by the wind at the rate:

\begin{equation}
{\dot M}_{\rm wind} = {\dot M}_{\rm sg} - {\dot M}_{\rm cr,2} .
\end{equation}

To calculate re-accretion of material from the wind by the
secondary, we assume that, as in eq. (5), the mass of
re-accreted material is proportional to the mass loss rate from
the white dwarf (${\dot M}_{\rm wind} $).

We may estimate the characteristic
time scales for each of angular momentum loss mechanisms during a strong wind
phase. We find, for example, that such a wind occurs for
a binary system with component masses 1.57 and
0.95 M$_\odot$, and an orbital period of 1.1~d.  In such a system, the
secondary has a radius of $\sim$2.6 ${\rm R}_\odot$, $f_1$=0.4, the
mass transfer rate from the secondary to white dwarf is
1.05$\times 10^{-6} ~{\rm M}_\odot ~{\rm yr}^{-1} $, $\dot{M}_{\rm wind} = 4
\times 10^{-7} ~{\rm M}_\odot ~{\rm yr}^{-1} $, and U=10.  The characteristic
time scales are then: $\tau_{\rm MSW} \sim 5 \times 10^{18}$ s,
$\tau_{\rm FAML} \sim 3 \times 10^{15}$ s and $\tau_{\rm FWIND} \sim 5
\times 10^{14}$ s.

\section{The evolutionary code}

Models of secondary stars filling their Roche lobes were computed
using a standard stellar evolution code based on the Henyey-type
code of Paczy\'nski (1970), which has been adapted to low-mass
stars (as described in detail in MS98). We use the Eggleton (1983)
formula to calculate the size of the secondary's Roche lobe.

For radiative transport, we use the opacity tables of Iglesias \&
Rogers (1996). Where the Iglesis \& Rogers (1996) tables are
incomplete, we have filled the gaps using the opacity tables of
Huebner et al. (1977). For temperatures lower than 6000 K, we use
the opacities given by Alexander \& Ferguson (1994) and Alexander
(private communication).

\subsection{Nuclear network}

Our nuclear reaction network is based on that of Kudryashov \&
Ergma (1980), who included the reactions of the CNO tri-cycle in
their calculations of hydrogen and helium burning in the envelope
of an accreting neutron star. We have included the reactions of
the proton-proton (PP) chain. Hence we are able to follow the
evolution of the elements: $^{1}$H, $^{3}$He, $^{4}$He, $^{7}$Be,
$^{12}$C, $^{13}$C, $^{13}$N, $^{14}$N, $^{15}$N, $^{14}$O,
$^{15}$O, $^{16}$O, $^{17}$O and $^{17}$F. We assume that the
abundances of $^{18}$O and $^{20}$Ne stay constant throughout the
evolution. We use the reaction rates of: Fowler, Caughlan \&
Zimmerman (1967, 1975), Harris at al. (1983), Caughlan et al.
(1985), Caughlan \& Fowler (1988), Bahcall \& Ulrich (1988),
Bahcall \& Pinsonneault (1992), Bahcall, Pinsonneault \&
Wesserburg (1995) and Pols et al. (1995).

\subsection{The role of unstable helium burning}

In recent years several papers concerning unstable helium burning
in accreting white dwarfs have been published. Cassini, Iben \&
Tornambe (1998) investigated accretion onto
cool C-O white dwarfs of masses 0.5 and 0.8 M$_\odot $. They
considered broad range of mass accretion rates from $10^{-8}$ to
$10^{-6} ~{\rm M}_\odot ~{\rm yr}^{-1} $. They found that even in the case
when hydrogen burns at the same rate at which it is accreting,
helium flash leads to envelope expansion. The expanded envelope,
due to interaction with the companion star is lost from the
system. The net result is that the mass of the accreting white
dwarf decreases.

Kato \& Hachisu (1999) proposed that if an optically thick wind
occurs during helium shell flashes, due to its high velocity
($\sim$1000 km s$^{-1}$) the wind leaves the system
without interaction with the companion star. They found that at a
relatively high accretion rate, the mass accumulation efficiency is
large enough for the white dwarf to grow above the Chandrasekhar
mass limit. However, it is important to note that they considered an
initially relatively massive (1.3 M$_\odot$) white dwarf, and
neglected restriction by hydrogen shell burning of the maximum
accretion rate (see formula (9)). Because of these approximations,
the mass accumulation ratio, $1 - f_1$, never equals unity.

Langer et al. (2000) considered the evolution of main sequence
star + white dwarf binary systems towards type Ia supernovae. They
assumed that above some critical accretion rate ($10^{-8} ~{\rm M}_\odot
~{\rm yr}^{-1} $ --- for Population I) helium burning proceeds such that
violent nova flashes and consequent mass ejection from the white
dwarf are avoided.

In our calculations, mass loss due to unstable helium burning has
been neglected. So that the masses of white dwarfs we obtain are
overestimated.

\section{Results of calculations}

\subsection{The grid of models}

To understand the evolution of close binary system consisting of
a C-O or NeMgO white dwarfs and near-main-sequence or
helium-rich stars, we computed various evolutionary
sequences for three  chemical compositions (Population II ---
X=0.756, Z=0.001; Population I --- X=0.68, Z=0.02; and helium rich
--- X=0.5, Z=0.02), for
 four  initial mass ratios $q_{\rm i} = M_{\rm sg,i}/M_{\rm wd,i}$ = 1.5, 2.0,
2.5, 3.0 and five white dwarf masses: 0.7, 0.85, 1.0, 1.15,
1.3 M$_\odot$. For each system  three different evolutionary
stages for secondary star are followed: i) the secondary fills its
Roche lobe as a M-S star with central hydrogen content $X_{\rm c} \sim
0.4 $; ii) the secondary fills its Roche lobe as a terminal M-S
star with small helium core $M_{\rm HeCore} \sim 0.01 ~{\rm M}_\odot$; and iii) the secondary
fills Roche lobe upon reaching helium core more $M_{\rm HeCore} \sim 0.1
~{\rm M}_\odot $ (Sarna, Ergma \& Ger\v{s}kevit\v{s}, in preparation). We found
that only a few evolutionary sequences are able to produce system
parameters similar to U Sco. Since the evolutionary stage of the
secondary (hydrogen- or helium-rich) is not observationally
well-determined, we selected from our grid of models four different
sequences with hydrogen and helium rich secondaries.  Initial
parameters of the sequences are presented in Table 1.

\begin{table}
\begin{center}
\begin{tabular}{lccccc}
\multicolumn{6}{l}{Table 1 ~~Initial parameters of computed sequences} \\
\hline
model  &  $M_{\rm sg} $ & $M_{\rm wd}$ & $P_{\rm i}$(RLOF) & $X$ & $Z$ \\
 & [$\rm M_\odot$] & [$\rm M_\odot$] & [d] & & \\
\hline
A        & 1.4 & 0.70 & 1.34 & 0.68 & 0.02 \\
B        & 1.7 & 0.85 & 1.61 & 0.68 & 0.02 \\
C        & 1.4 & 0.70 & 1.26 & 0.50 & 0.02 \\
D        & 1.7 & 0.85 & 1.24 & 0.50 & 0.02 \\
\hline
\end{tabular}
\end{center}
\end{table}

The sequences C and D represent the helium rich  channel proposed
by Hachisu et al. (1999). In Table 2  results for computed
sequences are shown. The masses of the subgiant and the white
dwarf are given at the moment when orbital period of the system is
equal to 1.23 d. The effective temperature and luminosity are for
the subgiant star.

An extended grid of multi-cycle nova evolution models by Prialnik
\& Kovetz (1995) allows to find entire range of observed nova
characteristics including recurrent and symbiotic novae. In Fig. 1,
three characteristic lines are
drawn, according to the PK95 calculations and their classification
scheme. Fast, slow and symbiotic novae are located below line 1 (we
used the classification scheme of PK95). Between lines 1 and 2,
recurrent novae with recurrence periods longer than 10 yr but
shorter than 100 yr are located. Above line 2 but below line 3, the mass
accretion rate onto the white dwarf is higher than $\dot{M}_{\rm cr,1} $
and nova outbursts do not occur --- hydrogen burning is stable.
Above line 3, the stable hydrogen burning is accompanied
by a strong optically thick wind. According to our
calculations, all models go first through an short nova phase
(n1). After that, all systems  enter into the first stable
hydrogen shell burning stage (s1). Sequences A and B do not exhibit a
wind phase. After the stable burning stage, system A evolves into second
nova phase (n2), while system B evolves into a recurrent nova phase.
Sequences C and D have a wind phase (w) and a second stable hydrogen
shell burning stage (s2). After the second stable hydrogen burning stage,
sequence C evolves through recurrent novae phase (rn). Sequence D avoids a
second nova phase. For this system, the white dwarf mass exceeds
Chandrasekhar limit during stable hydrogen burning, and a supernova
explosion may occur. In Table 3 the duration of each phase is shown. The
duration of the stable hydrogen burning and wind phases varies among
different sequences from several hundred thousand to several million years.

\begin{figure}
\epsfverbosetrue
\begin{center}
\leavevmode \epsfxsize=15.cm \epsfbox{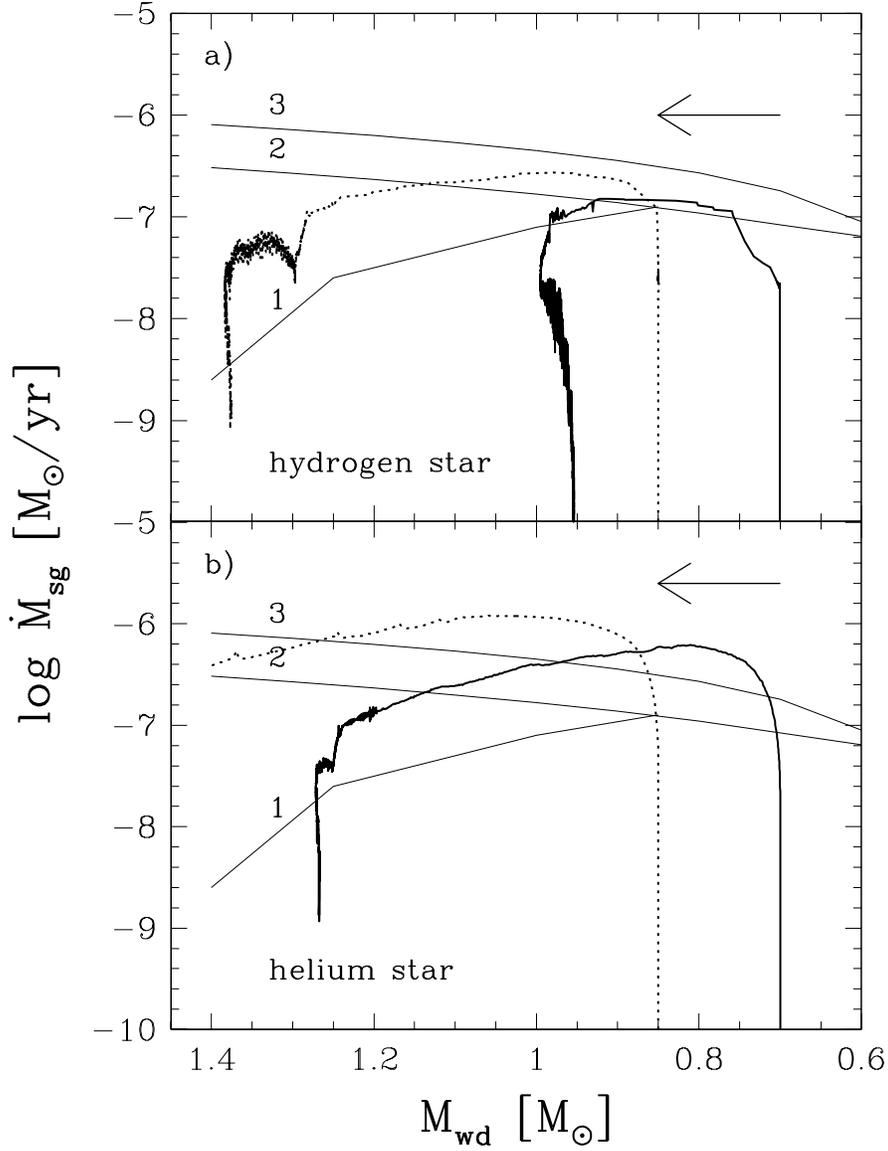}
\end{center}
\caption{The evolution of the mass transfer rate versus white
dwarf mass: (a) for a hydrogen-rich donor: sequence A (solid
line) and sequence B (dashed line) from Table 1; (b) for helium-rich
donor: sequence C (solid line) and sequence D (dashed line), also
from Table 1. The lines marked 1, 2 and 3 show lower boundaries for
recurrent novae, stable hydrogen burning and strong wind regions,
respectively. The region bounded by lines 1 and 2 shows the
recurrent nova phase. The arrows show the direction of evolution.}
\end{figure}

We infer that sequences B (hydrogen-rich) and C (helium-rich)
may evolve into systems like  U Sco.  After stable hydrogen
burning both sequences enter into a recurrent nova phase (Fig.1).

\begin{figure}
\epsfverbosetrue
\begin{center}
\leavevmode \epsfxsize=15.cm \epsfbox{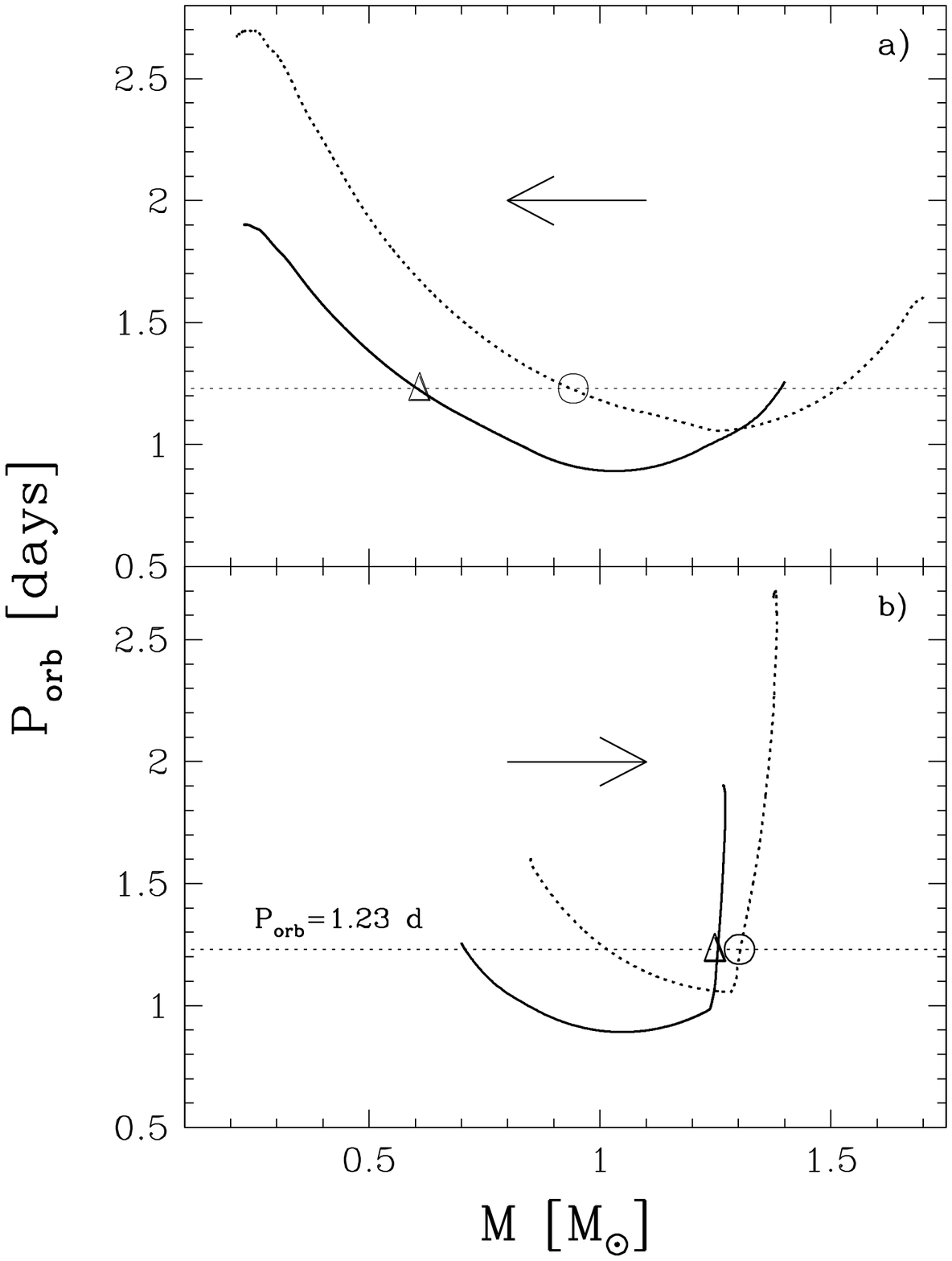}
\end{center}
\caption{The evolution of the orbital period of the system as a
function (a) of the subgiant mass $M_{\rm sg}$: solid line --- sequence
C, dashed line --- sequence B, and (b) of the white dwarf mass
$M_{\rm wd}$: solid line --- sequence C, dashed line --- sequence B.
Horizontal thin dashed lines mark the orbital period of U Sco. Open
circles mark positions of subgiant and white dwarf at this period for
sequence B, open triangles for sequence C. The arrows show the direction
of evolution.}

\end{figure}

For the same two evolutionary sequences, in Fig. 2 we present
evolution of the orbital period $P_{\rm orb}$ versus mass of the
subgiant $M_{\rm sg}$ (Fig. 2a) and mass of the white dwarf
$M_{\rm wd}$ (Fig. 2b). From our calculations we find that both
systems (B and C) evolve through the orbital period $P_{\rm orb} =
1.23$ d twice. During the first crossing of $P_{\rm orb} = 1.23$ d
line, the luminosity of the secondary is too high ($\log L/{\rm
L}_\odot \sim$ 0.9 and 1.35 for sequences B and C, respectively)
and it does not fit the observed absolute magnitude $M_{V}$=+3.8
of U Sco. During the second crossing, we find from the grid of
nova models (PK95 and KP97) outburst amplitudes $A = 7.6$ mag and
recurrence periods $P_{\rm rec}$ of 23 and 54 yrs, respectively,
for sequences B and C. The mean recurrence interval implied by
known outbursts is $P_{\rm rec} = 13$ yrs (excluding outburst in
1863). Therefore, we conclude that sequence B provides the best
fit to the observed parameters of U Sco.

\begin{table}
\begin{center}
\begin{tabular}{lccccc}
\multicolumn{6}{l}{Table 2 ~~Results for computed sequences} \\
\hline
model  &  $M_{\rm sg}$ & $M_{\rm wd}$ & $\log T_{\rm eff}$
 & $\log L/{\rm L}_\odot$ & $\dot{M}_{\rm sg}$ \\
 & [$\rm M_\odot $] & [$\rm M_\odot $] & [K] & & [$\rm M_\odot ~yr^{-1} $] \\
\hline A  &0.545   &0.983   &3.669  &0.153  &2.02$\times 10^{-8}$
\\ B  &0.936   &1.304   &3.693  &0.416  &3.58$\times 10^{-8}$ \\ C
&0.581   &1.263   &3.719  &0.370  &4.07$\times 10^{-8}$ \\ D
&1.696 &0.852   &3.957  &1.707  &1.14$\times 10^{-7}$ \\ \hline
\end{tabular}
\end{center}
\end{table}

\begin{table}
\begin{center}
\begin{tabular}{lcccccc}
\multicolumn{7}{l}{Table 3 ~~Time-scales for evolutionary phases}
\\ \hline model & $\Delta t_{\rm n1}$ & $\Delta t_{\rm s1} $ &
$\Delta t_{\rm w}$ & $\Delta t_{\rm s2}$ & $\Delta t_{\rm rn}$ &
$\Delta t_{\rm n2}$ \\ \multicolumn{7}{c}{[$\log (\Delta t/
{\rm yr})$]}
\\ \hline A &4.00 &6.57   &--     &--   &-- &8.17    \\ B   &5.97
&6.46 &-- &-- & 7.51 & 6.86    \\ C   &4.76     &5.41   &5.39
&6.17 & 7.17 & 6.72    \\ D &4.89     &5.02   &5.60   &5.58  &--
&--
\\ \hline
\end{tabular}
\end{center}
\end{table}

\subsection{Chemical composition of the subgiant and ejected matter}

From the grid of models calculated by PK95 and KP97, we can
interpolate the chemical composition of each model during its
nova phase. In Table 4, the isotopic compositions
in the envelope of the subgiant and in the ejected matter are shown
(at $P_{\rm orb} = 1.23$ d) for the two sequences B and C.

\begin{table*}
\begin{center}
\begin{tabular}{lcrrrcrcr}
\multicolumn{9}{l}{Table 4 ~~Chemical compositions (by mass) for sequences B
and C at $P_{\rm orb} = 1.23$ d.} \\ \hline model & $^{12}$C &
$^{13}$C & $^{14}$N & $^{15}$N  &
 $^{16} $O & $^{17}$O & $^{12}$C/$^{13}$C  &  $\rm N/C $ \\
& [$\rm \times 10^{-3} $] & [$\rm \times 10^{-4} $] & [$\rm \times
10^{-3} $] & [$\rm \times 10^{-7} $] & [$\rm \times 10^{-3} $] &
[$\rm \times 10^{-5} $] & & \\ \hline B subgiant & 1.41 & 3.05  &
3.16  & 5.64  & 9.77 & 0.58  & 4.6 & 1.8  \\ B ejecta   & 2.81 &
11.51 & 42.06 & 251.60 & 1.77 & 200.00 & 2.4 & 10.6 \\ C subgiant
& 0.31 & 1.07  & 4.69  & 2.91  & 9.77 & 1.10   & 2.9 & 11.3  \\ C
ejecta   & 1.93 & 7.10  & 35.89 & 94.82 & 3.82 & 51.75  & 2.7 &
13.6
\\ \hline
\end{tabular}
\end{center}
\end{table*}

If we compare the helium-rich model C with the hydrogen-rich model
B, we see that at $P_{\rm orb} = 1.23$ d the N/C ratio in the envelope
of the subgiant is higher for the helium-rich model,
but the isotopic ratio $^{12}$C/$^{13}$C is higher for the
hydrogen-rich model. In the ejected matter, both ratios are
similar for models B and C.

Our calculations show that as these models evolute, the He/H ratio of
the matter lost from the system changes from 0.56 to 1.26. For
sequences B and C, at $P_{\rm orb}$= 1.23 d the He/H ratio is 0.75 and
0.65, respectively, \emph{i.e.}, within the limits (0.4 -- 2) obtained from
observation (Anupama \& Dewangan 2000, Williams et al. 1981).

\begin{figure}
\epsfverbosetrue
\begin{center}
\leavevmode \epsfxsize=15.cm \epsfbox{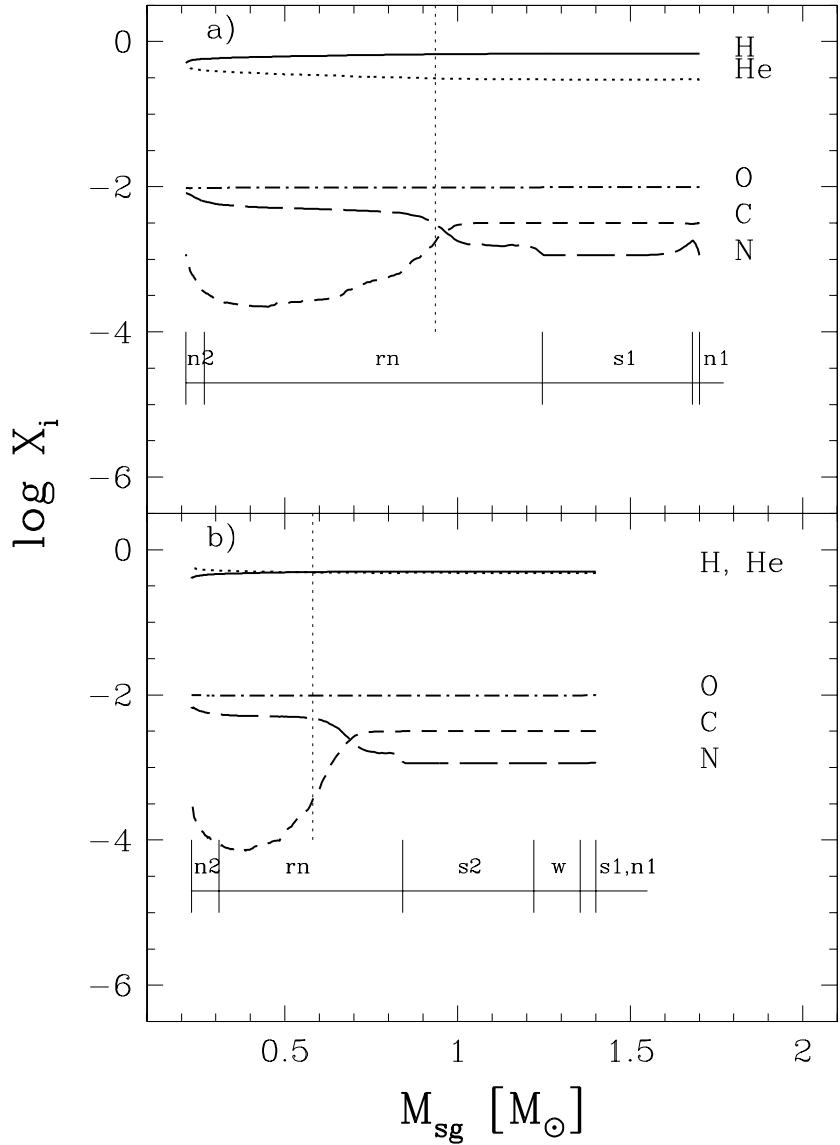}
\end{center}
\caption{The evolution of the red giant surface abundances (by mass) of H,
He, C, N and O as a function of the subgiant mass $M_{sg}$: upper
panel --- sequence B, lower panel --- sequence C. Vertical thin
dashed lines mark the position of subgiant mass for sequences B
($M_{\rm sg} = 0.926~{\rm M}_\odot$) and C ($M_{\rm sg} = 0.603 ~{\rm M}_\odot
$) at $P_{\rm orb} = 1.23$ d, respectively (see Table 2 for more
details). The first and second novae phases are marked n1 and n2,
respectively ; The first and second stable hydrogen shell
burning episodes are marked s1 and s2, respectively;
and the strong optically thick wind phase is marked w.
Systems evolve from right to left. For more explanation see text.}
\end{figure}

Figs. 3 and 4 show the evolution of the abundances by mass of helium,
hydrogen, carbon, nitrogen and oxygen of the subgiant envelope and
ejected matter. The vertical thin dashed lines show the place
where the orbital period is equal to 1.23 d. During stable
hydrogen shell burning phases, no loss of matter from the system
occurs. These phases are marked in Fig. 4 by setting $\log
X_{\rm i}=0$.

\begin{figure}
\epsfverbosetrue
\begin{center}
\leavevmode \epsfxsize=15.cm \epsfbox{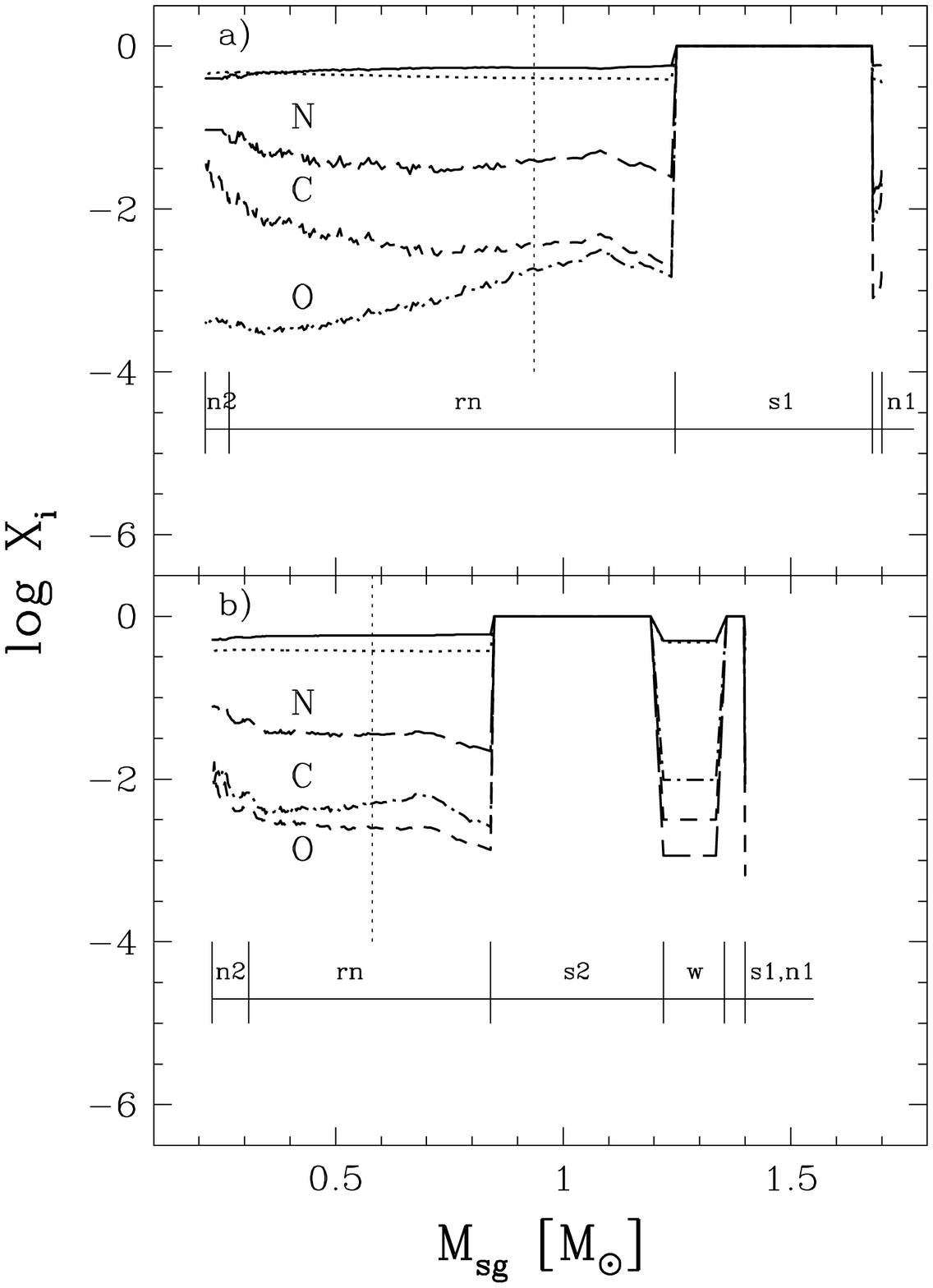}
\end{center}
\caption{The evolution of ejecta abundances (by mass) of H, He,
C, N and O as a function of the subgiant mass $M_{\rm sg}$: upper
panel --- sequence B, lower panel --- sequence C. Systems evolve from
right to left.  The various phases are the same as in Fig.3}
\end{figure}

\section{Observational tests}

Chemical composition and isotopic analysis may give more
information about the evolutionary stage of U Sco.  Unfortunately,
the subgiant component in U Sco is too faint for infrared
spectroscopic observations of CO bands in order to determine the
$^{12}$C/$^{13}$C ratio. However, we think that blue and red
domain spectra of U Sco could show some absorption structure in
the region of 4216\AA~and 7920\AA~in the CN sequence, similar to
that observed in DQ Her (Chanan, Nelson \& Margon 1978, Schneider
\& Greenstein 1979, Willimas 1983, Bauschlicher, Langhoff \&
Taylor 1988). Analysis of the blue region of the spectra is more
complicated because the structure of the CN violet system can be
affected by absorption features from the disc. However, we suggest
that since the matter in the accretion disc reflects the chemical
composition of the subgiant star, analysis of the disc will
provide the information we need if we use observations made during
the quiescent phase. We suggest that the red CN band is more
useful for observations. We propose to observe spectral region
near 7920--7940\AA~to identify two $^{13}$CN lines at
7921.13\AA~and 7935.67\AA~which are important for both
$^{12}$C/$^{13}$C isotopic ratio and N/C ratio determination. In
the case of low $^{12}$C/$^{13}$C ratio ($<$ 10), these lines are
clearly recognized (Fujita 1985), whereas in stars with
$^{12}$C/$^{13}$C $>$ 20 both lines are undetectable. Chemical
analysis of the expanding envelope (Anupama \& Dewangan 2000) also
would give useful information, allowing for comparison with
theoretical models (see Table 4).

\section{Discussion}

\subsection{Hydrogen-rich vs helium-rich models}

Hachisu et al. (1999) proposed a new evolutionary path to SNe Ia formation, in
which
the companion star is helium-rich. In their model, typical
orbital parameters of SNe Ia progenitors are:
$M_{\rm wd} = 1.37 ~{\rm M}_\odot$,
$M_{\rm sg} \sim  1.3 ~{\rm M}\odot$, and $\dot{M}_{\rm sg} \sim 2 \times
10^{-7} ~{\rm ~M}_\odot ~{\rm yr}^{-1}$. Based on light-curve analysis,
Hachisu et al. (2000) also constructed a detailed theoretical model for
U Sco. They found that the best fit
parameters are: $M_{\rm wd} \sim 1.37 ~{\rm M}_\odot$,
and $M_{\rm sg} \sim 1.5 ~{\rm M}_\odot$ (a range from 0.8 to 2.0 M$_\odot$
being acceptable).

However, the helium-enriched model poses a serious problem. Truran
et al. (1988) discussed the composition dependence of
thermonuclear runaway models for recurrent novae of U Sco-type.
They showed that for $M_{\rm wd} = 1.38 ~{\rm M}_\odot$, $\dot{M}_{\rm sg}
= 1.5 \times 10^{-8} ~{\rm M}_\odot ~{\rm yr}^{-1}$ and $L = 0.1 ~{\rm
L}_\odot$ optically bright outbursts are obtained only for matter
with He/H$<$1 by mass. Their results are consistent with our sequences
B and C where He/H are equal 0.47 and 0.98, respectively.
The large helium content observed in U Sco must then be attributed to
helium enrichment from the underlying white dwarf (Prialnik \& Livio
1995).

According to Kato (1996), a supersoft component in UV spectrum
is predicted to be observable about 10 days after the outburst.
For the hydrogen-rich model (He/H=0.1) the supersoft X-ray
component is expected to rise until $\sim 50$ days after the
outburst to a maximum luminosity of $\sim 3 \times
10^{36}$ erg s$^{-1}$ $(d/{\rm kpc})^2$. For the helium-rich model (He/H
= 2), the maximum luminosity is reached about 20 days after the
optical outburst.  The evolution of the X-ray
luminosity could therefore give important evidence about the chemical
composition of the accreted matter.  Unfortunately, only a synoptic soft
X-ray detection of U Sco exists, some 19--20 days after outburst (Kahabka et
al. 1999).

\subsection{The recurrent novae phase}

In MS98, a grid of evolutionary sequences for $q_{\rm i} = 1.25$ and
white dwarf initial masses from 0.8 to 1.2 M$_\odot$ has been
calculated. According to those calculations it is possible to
produce a system like U Sco (for Population I chemical composition),
but only if the initial white dwarf mass is already very high ($>
1.2 ~{\rm M}_\odot$). For the model with $M_{\rm wd,i} = 1.2 ~{\rm M}_\odot$,
the system spends less than 1\% of its semidetached evolution ($5 \times
10^8$ yr) in the recurrent novae phase. For the rest of the time, the
system evolves through fast novae and moderately  fast novae
phases. Our recent calculations for $q_{\rm i} =1.25$ show that systems
with $M_{\rm wd,i} = 1.3 ~{\rm M}_\odot$ evolve towards the Chandrasekhar
limit (Supernovae Ia). The system spends about 30\% of its
semidetached evolution ($4.2 \times 10^8$ yr) in the recurrent nova
phase. In the calculations described here, which begin with higher initial
mass ratios and lower-mass white dwarfs, sequences B
(hydrogen-rich) and C (helium-rich) both evolve through the recurrent
novae phase, and spend in this phase 75\% and 65\%of their
semidetached evolution, respectively (see Table 3). In their final phases
of semidetached evolution, massive white dwarfs near the
Chandrasekhar limit are formed, but they never exceed this limit.

\section{Conclusions}

We calculated several evolutionary sequences to reproduce the orbital
and physical parameters of the recurrent novae of U Sco-type. The
results of these calculations can be summarized as follows:

(1) We find a new evolutionary channel for the formation of the U
Sco-type systems. We show that they may form from
binaries with a low-mass C-O white dwarf as accretor if
$M_{\rm wd,i}< 0.85 ~{\rm M}_\odot$ and $1.5 < q_{\rm i} < 2.5$. Such systems evolve
through several observable stages: SSS with stable hydrogen
burning, strong wind phase, and long-term recurrent nova phase
(longer than $10^7$ years, Table 3). In the final phase of evolution,
massive white dwarfs near the Chandrasekhar mass limit are formed,
but they never exceed this limit.

(2) We find that our evolutionary sequence B is able to produce a binary
system with parameters similar to U Sco.
This model has initial parameters: $M_{\rm sg,i} = 1.7 ~{\rm M}_\odot $,
$M_{\rm wd,i} = 0.85 ~{\rm M}_\odot$ and $P_{\rm i} {\rm (RLOF)} = 1.61$ d. Based on this
evolutionary model, we found that
the best fit parameters for U Sco are: $M_{\rm sg} = 0.94 ~{\rm M}_\odot $, $M_{\rm wd} =
1.31~{\rm M}_\odot$, $\log L_{\rm sg}/{\rm L}_\odot = 0.42 $, and $\dot M_{\rm sg} = 3.58 \times
10^{-8} ~{\rm M}_\odot ~{\rm yr}^{-1}$ for $P_{\rm orb} = 1.23056$ d.

\section*{\sc Acknowledgments}

This work is partly supported through
grants 1~P03D 024 26 from the Ministry of Scientific Research and
Informational Technology, Poland {\bf and Collaborative Linkage
Grant PST.CLG.977383}. JG and EE acknowledge support through
Estonian SF grant 4338. EE acknowledges the warm hospitality of
the Astronomical Institute ``Anton Pannekoek'' where part of this
work was conducted. While in Netherlands, EE was supported by NWO
Spinoza grant 08-0 to E. P. J. van den Heuvel.

\end{document}